\documentclass[a4paper,12pt]{article}
\pdfoutput=1
\usepackage{amssymb}
\usepackage{epsfig}
\usepackage{graphicx}
\usepackage{slashed}

\usepackage[utf8]{inputenc}

\makeatletter

\@addtoreset{equation}{section}
\makeatother
\def\be{\begin{equation}}
\def\ee{\end{equation}}
\def\bea{\begin{eqnarray}}
\def\eea{\end{eqnarray}}

\def\({\left(}
\def\){\right)}
\def\<{\left<}
\def\>{\right>}

\def\be{\begin{equation}}
\def\ee{\end{equation}}
\def\bea{\begin{eqnarray*}}
\def\eea{\end{eqnarray*}}
\def\ben{\begin{eqnarray}}
\def\een{\end{eqnarray}}
\def\({\left(}
\def\){\right)}
\def\<{\left<}
\def\>{\right>}
\def\!{\right|}
\def\|{\left|}

\def\[{\left[}
\def\]{\right]}

\def\+{\bar}
\def\mb{\mathbb}

\def\D{{\cal{D}}}
\def\L{{\cal{L}}}
\def\t{\widetilde}

\def\C{{\cal{C}}}

\def\N{{\cal{N}}}

\def\F{{\cal{F}}}

\def\W{{\cal{W}}}

\def\P{{\cal{P}}}
\def\L{{\cal{L}}}
\def\D{{\cal{D}}}

\def\K{{\cal{K}}}

\def\L{{\cal{L}}}

\def\eps{{\cal{\varepsilon}}}

\def\E{{\cal{E}}}

\def\F{{\cal{F}}}

\def\h{\widehat}

\begin{document}

\setlength{\unitlength}{1mm}

\pagestyle{empty}
\vskip-10pt
\vskip-10pt
\hfill 
\begin{center}
\vskip 3truecm
{\Large \bf
Five-dimensional Super-Yang-Mills
\vskip 0.3truecm
and its Kaluza-Klein tower}
\vskip 2truecm
{\large \bf
Andreas Gustavsson}
\vspace{1cm} 
\begin{center} 
Physics Department, University of Seoul, 13 Siripdae, Seoul 130-743 Korea
\end{center}
\vskip 0.7truecm
\begin{center}
(\tt agbrev@gmail.com)
\end{center}
\end{center}
\vskip 2truecm
{\abstract{We compactify the abelian 6d (1,0) tensor multiplet on a circle bundle, thus reducing the theory down to 5d SYM while keeping all the KK modes. This abelian classical field theory, when interpreted suitably, has a nonlocal superconformal symmetry. Furthermore, a nonabelian generalization, where all the KK modes are kept, is possible for the nonlocal superconformal symmetry, whereas for the local superconformal symmetry we can only realize a subgroup.}}
\vfill
\vskip4pt
\eject
\pagestyle{plain}

\section{Introduction}
One proposal for the 6d $(2,0)$ tensor multiplet compactified on a circle is that this is fully captured by the dimensionally reduced 5d maximally SYM \cite{Douglas:2010iu}, \cite{Lambert:2010iw}, where instanton particles are believed to play the role of all the KK particles. For abelian gauge group, a precise match between all instanton particles and all KK particles was found in \cite{Kim:2011mv}, \cite{Bak:2012ct}.

Another proposal has been to add a KK tower of fundamental field excitations to the 5d SYM. It has been shown that this can be done while preserving the 6d $(1,0)$ Poincare supersymmetry in flat space with one compact circle direction \cite{Bonetti:2012st}, \cite{Ho:2011ni}, \cite{Ho:2014eoa}, \cite{Gustavsson:2018pov}. In this paper, we will generalize this construction and consider six-manifolds that are circle-bundles. On such manifolds, we will realize the 6d $(1,0)$ superconformal symmetry $\C_{6d,(1,0)}$ on 5d SYM plus the KK tower. But not in a conventional way. This symmetry acts in a nonlocal way on the fields. Why we need a nonlocal variation is easy to understand. We assume the existence of a supersymmetry parameter that solves the 6d conformal Killing spinor equation
\ben
D_M \eps &=& \Gamma_M \eta\label{6dCKE}
\een
Such a supersymmetry parameter depends in general on the location along the circle fiber. If we use this parameter to make a local variation of a zero mode field, that is, a field in the 5d SYM multiplet, then the variation of that field has to involve higher KK modes. If on the other hand we vary the 5d SYM field nonlocally, then we can do that without bringing in the KK modes into the variation. Thus with a nonlocal variation we can vary the zero mode fields in a closed way among themselves despite the supersymmetry parameter itself is not a zero mode. If we restrict ourselves to a local symmetry, then we can still realize a 5d restriction $\C_{5d,(1,0)}$ of $\C_{6d,(1,0)}$, where the supersymmetry parameter, in addition to satisfying (\ref{6dCKE}), is constant along the circle fiber,
\ben
\partial_t \eps(t) &=& 0\label{5d}
\een
Here we parametrize the position along the fiber by the time coordinate $t$. The time derivative coincides with the Lie derivative along the fiber since we use the standard circle bundle metric, and a reparametrization of $t$ is not allowed because that will take us outside the standard circle bundle form for the metric. Alternatively, we may write the condition (\ref{5d}) in a coordinate independent way as
\bea
\L_V \eps &=& 0
\eea
where $V$ is the Killing vector field along the circle fiber. In \cite{Gustavsson:2018rcc} we showed that 5d SYM has the classical symmetry $\C_{5d,(1,0)}$.

\section{The six-sphere}
To show the existence of a solution to both (\ref{6dCKE}) and (\ref{5d}) on a curved space, it seems that the round $S^6$ will be the easiest example to study. The metric on $S^6$ with radius $r$ in polar coordinates is given by
\bea
ds^2_{S^6} &=& r^2 \(d\theta^2 + \sin^2 \theta ds^2_{S^5}\)
\eea
where
\bea
ds^2_{S^5} &=& \(d\tau + \kappa_i dx^i\)^2 + ds_{\mb{CP}^2}^2
\eea
is the metric on the equatorial $S^5$, viewed as a circle-bundle over $\mb{CP}^2$ with fiber coordinate $\tau\sim \tau+2\pi$ and graviphoton $\kappa_i$ whose field strength $w_{ij} = \partial_i \kappa_j - \partial_j \kappa_i$ is proportional to the Kahler two-form on $\mb{CP}^2$. By making the coordinate transformation
\bea
R &=& 2 r \tan \frac{\theta}{2}
\eea
we get
\bea
ds^2_{S^6} &=& e^{2\sigma} \(dR^2 + R^2 ds_{S^5}^2\)\cr
e^{\sigma(R)} &=& \frac{1}{1+\frac{R^2}{4r^2}}
\eea
This shows that $S^6$ is conformally flat. We may write this metric in the standard circle-bundle form
\ben
ds^2_{S^6} &=& e^{2\sigma(R)} R^2 \(d\tau + \kappa_i dx^i\)^2 + e^{2\sigma(R)} \(dR^2 + R^2 ds^2_{\mb{CP}^2}\)\label{S6metric}
\een 
We may also express this same metric as 
\bea
ds^2_{S^6} &=& e^{2\sigma} dx^M dx^M\cr
e^{\sigma} &=& \frac{1}{1+\frac{x^M x^M}{4r^2}}
\eea
where $x^M$ are Euclidean coordinates on $\(\mb{R}^6,\delta_{MN}\)$. In terms of these coordinates, the most general solution to (\ref{6dCKE}) is given by \cite{Naseer:2018cpj}
\bea
\eps &=& e^{\frac{\sigma}{2}} \(\eps_1 + x^M \Gamma_M \eps_2\)
\eea
where $\eps_1$ and $\eps_2$ are constant parameters, $\partial_M \eps_{1,2} = 0$. The Killing spinor solution 
\bea
\eps_0 &=& e^{\frac{\sigma(R)}{2}} \eps_1
\eea
survives the dimensional reduction along the Hopf fiber along the $\tau$-direction since it does not depend on the coordinate $\tau$. One may show that once a metric is in the standard circle-bundle form, the Lie derivative along the $\tau$ direction is simply given by $\L_{\frac{\partial}{\partial_\tau}} \eps = \partial_{\tau} \eps$. We note that $\sigma$ depends on the coordinate $R = \sqrt{x^M x^M}$ in the metric (\ref{S6metric}), but it does not depend on $\tau$. Thus we have now showed the existence of such a supersymmetry parameter on the curved space $S^6$.

\section{Superconformal algebra in curved space}
Let us start with a realization of the $\C_{6d,(2,0)}$ symmetry. We assume an abelian gauge group. The supersymmetry parameter $\eps$ satisfies the conformal Killing spinor equation (\ref{6dCKE}) and the 6d Weyl projection $\Gamma \eps = -\eps$. The $(2,0)$ tensor multiplet consists of a Weyl fermion of opposite chirality $\Gamma \Psi = \Psi$, five scalar fields $\phi^A$ and a two-form gauge field $B_{MN}$. The supersymmetry variations are 
\ben
\delta \phi^A &=& i \bar\eps \tau^A \Psi\cr
\delta B_{MN} &=& i \bar\eps \Gamma_{MN}\psi\cr
\delta \Psi &=& \frac{1}{12} \Gamma^{MNP} \eps H_{MNP} - \Gamma^{M} \tau^A \eps \partial_M \phi^A - 4 \tau^A \eta \phi^A\label{6dtensor}
\een
The supersymmetric Lagrangian is given by
\bea
\L &=& \frac{1}{2\pi} \(-\frac{1}{24} H_{MNP}^2 - \frac{1}{2} (D_M\phi^A)^2 - \frac{R}{10} (\phi^A)^2 + \frac{i}{2} \bar\Psi\Gamma^M D_M\Psi\)
\eea
More precisely, its supersymmetry variation is given by a total derivative
\bea
\delta \L &=& D_M \(-\frac{i}{4\pi}\bar\Psi \(\frac{1}{12}\Gamma^{RST}\Gamma^M\eps H_{RST} - \Gamma^R \tau^A \Gamma^M \eps D_R \phi^A - 4 \tau^A \Gamma^M \eta \phi^A\)\)
\eea
If we assume the six-manifold is closed, then the action will be supersymmetric on any six-manifold where (\ref{6dCKE}) has some solution. For instance on $S^6$. However, the above Lagrangian is written in Lorentzian signature. 

In a curved space, we shall distinguish between the Weyl weight $\W$ and the scaling dimension $\Delta$. If the field has tensor indices in the spacetime directions, then these indices contribute to the scaling dimension but not to the Weyl weight. For instance, $B_{MN}$ has Weyl weight $\W = 0$ and scaling dimension $\Delta = 2$. One way to see this is by looking at the action for a nonchiral tensor gauge field with field strenght $H_{MNP}=(dB)_{MNP}$ in curved space, 
\bea
- \frac{1}{12} \int d^6 x \sqrt{-g} g^{MM'} g^{NN'} g^{PP'} H_{MNP} H_{M'N'P'}
\eea
This is invariant under the Weyl transformation
\bea
g'_{MN} &=& e^{2\Omega} g_{MN}\cr
B'_{MN} &=& B_{MN}
\eea
Since $B_{MN}$ is not Weyl rescaled, we will say that it has Weyl weight zero, $\W = 0$. On the other hand, $B_{MN}$ has scaling dimension $\Delta = 2$. The scaling dimension appears in the superconformal algebra in flat space. It does not appear in the superconformal algebra in curved space. What appears in the superconformal algebra in curved space, is the Weyl weight. 

In curved space, the closure relations are
\bea
\delta^2 \phi^A &=& - i \L_S \phi^A - 2 i \W_{\phi} \bar\eps\eta \phi^A - 4 i \bar\eps\tau^{AB} \eta \phi^B\cr
\delta^2 B_{MN} &=& - i \L_S B_{MN} - 2 i \W_B \bar\eps\eta B_{MN} + 2 \partial_M \Lambda_N\cr
\delta^2 \Psi &=& - i \L_S \Psi - 2 i \W_{\psi} \bar\eps\eta \Psi - i \bar\eps\tau^{AB}\eta \tau^{AB} \Psi\cr
&& + \frac{3i}{8} S_N \Gamma^N \Gamma^M D_M\Psi\cr
&& - \frac{i}{8} \bar\eps\Gamma_N\tau^A\eps\Gamma^N \tau^A \Gamma^M D_M\Psi
\eea
where $S^M = \bar\eps\Gamma^M \eps$ is a conformal Killing vector, satisfying 
\bea
D_M S_N + D_N S_M &=& \frac{1}{3} g_{MN} D^P S_P
\eea
Our spinor conventions are summarized in the appendix. The gauge parameters are
\bea
\Lambda_M &=& - i B_{MN}S^N+i\bar\eps\Gamma_M\tau^A\eps\phi^A
\eea
and the Weyl weights are
\bea
\W_{\phi} &=& 2\cr
\W_B &=& 0\cr
\W_{\psi} &=& \frac{5}{2}
\eea
From the above closure relations, we see that the superconformal algebra in curved space should contain the relation
\ben
\delta^2 &=& - i \L_S - 2 i \W \bar\eps\eta - 2 \bar\eps\tau^{AB}\eta S^{AB}\label{curvedSCA}
\een
where the generator of $SO(5)$ R-symmetry is represented as
\bea
(S^{AB})^{CD} &=& 2 i \delta^{AB,CD}\cr
S^{AB} &=& \frac{i}{2} \tau^{AB}
\eea
We can recover from (\ref{curvedSCA}) the usual superconformal algebra relation on flat $\mb{R}^{5,1}$ where the most general solution to (\ref{6dCKE}) is given by
\bea
\eps &=& \eps_0 + \Gamma_M \eta x^M
\eea
All the conformal transformations are encoded in the Lie derivative, and in flat space this Lie derivative becomes
\bea
S^M \P_M &=& \bar\eps_0 \Gamma^M \eps_0 \P_M + 2 \bar\eps_0 \eta \D + 2 \bar\eps_0 \Gamma^{MN} \eta \L_{MN} - \bar\eta \Gamma^M \eta \K_M
\eea
where
\bea
\P_M &=& - i \partial_M\cr
\D &=& - i x^M \partial_M\cr
\L_{MN} &=& i \(x_M \partial_N - x_N \partial_M\)\cr
\K_M &=& - 2 x_M x^N \partial_N + |x|^2 \partial_M
\eea
and (\ref{curvedSCA}) leads to 
\bea
\delta^2 &=& \bar\eps_0\Gamma^M\eps_0 P_M + 2 \bar\eps_0\eta D + 2 \bar\eps_0\Gamma^{MN}\eta L_{MN} - \bar\eta\Gamma^M\eta K_M + 2 \bar\eps_0 \tau^{AB}\eta S^{AB} 
\eea
which is part of the $(2,0)$ superconformal algebra on flat space $\mb{R}^{1,5}$, where
\bea
P_M &=& \P_M\cr
D &=& \D - i \Delta\cr
L_{MN} &=& \L_{MN} + S_{MN}\cr
K_M &=& \K_M - 2 i \Delta x_M - S_{MN} x^N
\eea
It is particularly interesting to note how the Lorentz generators $L_{MN}$ appear on the right-hand side in this superconformal algebra. These Lorentz generators do not appear if we limit ourselves to the Poincare supercharges that gives us only the translational symmetries. 

The advantage with working with the superconformal algebra on curved space is that we may keep a compact circle direction in the manifold and yet we may ask whether the theory has the superconformal symmetry. By leaving flat space, we may avoid the difficult question of how to take the decompactification limit of that circle and the question of Lorentz symmetry in that limit.

If we consider only the supersymmetry parameter $\eps_0 = e^{\sigma/2} \eps_1$ on $S^6$ and conformally map this to $\mb{R}^6$, then we get just the Poincare supersymmetries on $\mb{R}^6$ and this can only show us the translational symmetries. To see the Lorentz symmetries on $\mb{R}^6$ we need the full 6d superconformal algebra and for that we need to show the existence of all the higher KK towers of supercharges as well. On $S^6$, those higher supercharges are corresponding to the other solution $\eps = e^{\sigma/2} x^M \Gamma_M \eps_2$.

\section{Dictionary between 6d and 5d languages}
We will assume the 6d metric is of the general circle-bundle form 
\bea
ds^2 &=& - r^2 (dt + \kappa_m dx^m)^2 + G_{mn} dx^m dx^n
\eea
We take time to be compact $t \sim t + 2\pi$ and we define
\bea
w_{mn} &=& \partial_m \kappa_n - \partial_n \kappa_m
\eea
To get a 5d formulation, we expand all the 6d fields in terms of modes 
\bea
\Phi(t,x^m) &=& \Phi_0(x^m) + \sum_{n\in \mb{Z}} \Phi_n(x^m) e^{i n t}
\eea
To get to the 6d theory on the Euclidean space whose metric is 
\bea
ds^2 &=& r^2 (dt + \kappa_m dx^m)^2 + G_{mn} dx^m dx^n
\eea
what we need to do from the 5d perspective, is to replace the KK mode number $n$ everywhere by $i n$, but we will not do this here. 

The 6d conformal Killing spinor equation expressed in terms of 5d quantities reads \cite{Linander:2011jy}
\ben
\t D_m \eps &=& \gamma_m \eta - \frac{r}{4} w_{mn} \gamma^n \eps + \kappa_m \partial_0\eps\label{5dCKEi}\cr
\eta &=& \frac{1}{r}\partial_0\eps - \frac{r}{8} w_{mn}\gamma^{mn}\eps + \frac{1}{2r} (\partial_m r) \gamma^m \eps\label{5dCKEii}
\een
Here we put a tilde on 5d quantities, so for instance $\t D_m$ is the 5d spinor derivative that uses the 5d spin connection. The precise expressions for the derivatives acting on spinors are presented in the appendix.

As we want closure on a 6d Lie derivative but we use a 5d formulation, we first need to understand how to express the 6d Lie derivatives in terms of 5d quantities. First we relate the 6d quantity
\bea
S^M &=& \bar\eps\Gamma^M \eps
\eea
with the 5d quantities
\bea
s^m &=& \bar\eps\gamma^m\eps\cr
s &=& -\bar\eps\eps
\eea
By using corresponding relations between the 6d and 5d gamma matrices, we get the following relations
\bea
S^m &=& s^m\cr
S^0 &=& - \frac{s}{r} - \kappa_m s^m
\eea
and
\bea
S_m &=& s_m + \kappa_m r s\cr
S_0 &=& r s
\eea
Let us now consider the 6d Lie derivatives on a scalar, a two-form and a spinor,
\bea
\L_S \sigma &=& S^M \partial_M \sigma\cr
\L_S B_{MN} &=& S^P \partial_P B_{MN} + (\partial_M S^P) B_{PN} + (\partial_M S^P) B_{MP}\cr
\L_S \chi &=& S^M D_M \chi + \frac{1}{4} (D_M S_N) \Gamma^{MN} \chi
\eea
If we express these 6d Lie derivatives in terms of 5d quantities, they become
\bea
\L_S \sigma &=& s^m \(\partial_m \sigma - \kappa_m \partial_0 \sigma\) - \frac{s}{r} \partial_0 \sigma\cr
\L_S A_m &=& s^p \(\partial_p A_m - \kappa_p \partial_0 A_m\) - \frac{s}{r} \partial_0 A_m + (\partial_m s^p) A_p \cr
&& - \frac{1}{r} (\partial_0 s) A_m - \kappa_n (\partial_0 s^n) A_m - (\partial_0 s^n) B_{nm}\cr
\L_S B_{mn} &=& s^p \(\partial_p B_{mn} - \kappa_p \partial_0 B_{mn}\) + 2 (\partial_{[m} s^p) B_{|p|n]} - \frac{s}{r} \partial_0 B_{mn}\cr
&& + 2 \partial_m \(\frac{s}{r} + \kappa_q s^q\) A_n\cr
\L_S \chi &=& s^m (\t D_m \chi - \kappa_m \partial_0 \chi) - \frac{s}{r} \partial_0 \chi + \frac{1}{4} \(\partial_m s_n + r s w_{mn}\) \gamma^{mn} \chi
\eea
Alternatively we can write
\bea
\L_S A_m &=& s^p \partial_p A_m + (\partial_m s^p) A_p \cr
&& - \partial_0 \(s^p\kappa_p A_m+\frac{s}{r} A_m\) - (\partial_0 s^n) B_{nm}
\eea
Let us also here note the following identities,
\bea
\partial_m s &=& r w_{mn} s^n + \frac{s}{r} \partial_m r - \frac{1}{r} \partial_0 s_m + \kappa_m \partial_0 s\cr
\partial_m (r s) + 4 r \bar\eps\gamma_m\eta &=& \partial_0 s_m + r \kappa_m \partial_0 s\cr
\partial_m s_n + 2 \bar\eps\gamma_{mn}\eta &=& \frac{r s}{2} w_{mn} + \kappa_m \partial_0 s_n
\eea
The abelian 6d fermionic equation of motion
\bea
\Gamma^M D_M \chi &=& 0
\eea
becomes in the 5d language
\bea
\gamma^m (\t D_m \chi - \partial_0 \kappa_m \chi) - \frac{1}{r} \partial_0\chi + \frac{1}{2r} (\partial_m r) \gamma^m \chi - \frac{r}{8} w_{mn} \gamma^{mn} \chi &=& 0
\eea

\section{The $(1,0)$ supermultiplets}
As was noted in \cite{Bonetti:2012st}, it is necessary to break supersymmetry by half in order to write down a non-abelian generalization. We impose the Weyl projection condition
\bea
\tau^5 \eps &=& - \eps
\eea
thus breaking $SO(5)$ R-symmetry down to $SO(4) = SU(2)_R \times SU(2)_F$, where the first factor $SU(2)_R$ is the resulting R symmetry and the second factor $SU(2)_F$ is a flavor symmetry. We then use the index notation $A=(i,5)$ where $i=1,2,3,4$. This will reduce the amount of supersymmetry by half, to $(1,0)$, and it splits the $(2,0)$ tensor multiplet fermion into a $(1,0)$ tensor multiplet fermion $\chi$ and a hypermultiplet fermion $\zeta$ subject to 
\bea
\tau^5 \chi &=& -\chi\cr
\tau^5 \zeta &=& \zeta
\eea
The supersymmetry variations for the abelian $(1,0)$ tensor multiplet in the 5d langauge were obtained in the appendix in \cite{Gustavsson:2018pov}. They are given by
\bea
\delta \sigma &=& - i \bar\eps \chi\cr
\delta A_m &=& i r \bar\eps \gamma_m \chi\cr
\delta B_{mn} &=& i \bar\eps \gamma_{mn} \chi - 2 i r \kappa_{[m} \bar\eps\gamma_n \chi\cr
\delta \chi &=& \frac{1}{2r} \gamma^{mn} \eps \F_{mn} + \gamma^m \eps  \(\partial_m\sigma - in\kappa_m \sigma\) + \frac{in}{r}\eps\zeta + 4\eta\sigma
\eea
For the hypermultiplet one may likewise obtain
\bea
\delta \sigma^i &=& i \bar\eps \tau^i \zeta\cr
\delta \zeta &=& - \gamma^m \tau^i \eps  \(\partial_m\sigma^i - in\kappa_m \sigma^i\) - \frac{in}{r}\tau^i\eps\sigma^i - 4\tau^i\eta\sigma^i
\eea
All our fields carry a hidden KK mode index $n$ in terms of which we represent $\partial_0$ as $in$ where $n$ is integer. For the abelian case we do not need a separate treatment of the zero modes $n=0$ but the above variations apply to those zero modes as well. Now since these are nothing but a rewriting of the 6d supersymmery variations, there must be the infinite tower of KK supercharges satisfying (\ref{5dCKEi}).

\section{Abelian supersymmetry}
As we will introduce a more general mode expansion below, let us now write the 6d supersymmetry variations for the tensor multiplet without any mode expansion where we use the time derivative $\partial_0$ instead of writing it in terms of the modes. Then we have
\bea
\delta \sigma &=& - i \bar\eps \chi\cr
\delta A_m &=& i r \bar\eps \gamma_m \chi\cr
\delta B_{mn} &=& i \bar\eps \gamma_{mn} \chi - 2 i r \kappa_m \bar\eps\gamma_n\chi\cr
\delta \chi &=& \frac{1}{2r} \gamma^{mn} \eps \F_{mn} + \gamma^m \eps \(D_m \sigma - \kappa_m \partial_0 \chi\) + \frac{1}{r} \eps \partial_0 \sigma + 4 \eta \sigma
\eea
The closure relations are
\bea
\delta^2 \sigma &=& - i \L_S \sigma - 4 i \bar\eps\eta \sigma\cr
\delta^2 A_m &=& - i \L_S A_m + \(\partial_m \Lambda_0 - \kappa_m \partial_0 \Lambda_0\) - \partial_0 \t\Lambda_m\cr
\delta^2 B_{mn} &=& - i \L_S B_{mn} + 2 \partial_m \(\t\Lambda_n + \kappa_m \Lambda_0\)\cr
&& + i s^r \E[B]_{rmn}\cr
\delta^2 \chi &=& - i \L_S \chi - 5 i \bar\eps\eta \chi + \frac{i r}{8} s^{mn,ij} w_{mn} \tau^{ij} \chi\cr
&& + \(- \frac{3 i}{8} s+ \frac{3 i}{8} s_p \gamma^p + \frac{i}{64} s^{ij}_{pq} \gamma^{pq} \tau^{ij}\)\E[\chi]
\eea
We thus have closure on-shell, that is when we put
\bea
\E[B]_{rmn} &=& H_{rmn} + \frac{1}{2r} \E_{rmn}{}^{pq} \F_{pq} - 3 \kappa_r \F_{mn}\cr
\E[\chi] &=& \gamma^m (\t D_m \chi - \kappa_m \partial_0\chi) - \frac{1}{r} \partial_0 \chi + \frac{1}{2r} (\partial_m r) \gamma^m \chi - \frac{r}{8} w_{mn} \gamma^{mn} \chi 
\eea
to zero. We then have closure up to a gerbe gauge transformation, with the gauge parameters
\bea
\Lambda_0 &=& i \(s^m A_m - r s \sigma\)\cr
\t\Lambda_m &=& \frac{i}{r} \(s A_m - r s_m \sigma\) + i s^n \(\kappa_n A_m - \kappa_m A_n\) - i B_{mn} s^n
\eea

The superconformal Lagrangian is given by \cite{Lee:2000kc}, \cite{Linander:2011jy}, \cite{Bonetti:2012fn}, \cite{Bonetti:2012st}, \cite{Cordova:2013bea} 
\ben
\L &=& \frac{1}{4r} \F^{mn} \F_{mn} + \frac{1}{8} \E^{mnpqr} \F_{mn}\partial_0^{-1}\(\partial_r-\kappa_r\partial_0\)\F_{pq} \kappa_r  \cr
&&+ \frac{1}{2r^3} (\partial_0\sigma)^2 - \frac{1}{2r} G^{mn} \(\partial_m \sigma-\kappa_m\partial_0\sigma\)\(\partial_n\sigma-\kappa_n\partial_0\sigma\) - \frac{K}{2r} \sigma^2 \cr
&& + \frac{i}{2r^2} \bar\chi \partial_0\chi - \frac{i}{2r} \bar\chi\gamma^m \(\t\D_m\chi-\kappa_m\partial_0\chi\) + \frac{i}{16r} \bar\chi \gamma^{mn} \chi w_{mn}\label{abelL}
\een
where ($\t R$ denotes the Ricci scalar on the base manifold)
\bea
K &=& \frac{\t R}{5} - \frac{r^2}{20} w_{mn}^2 - \frac{3}{5r} D^2_m r
\eea
Here we introduce $\partial_0^{-1}$ that we shall define as the inverse of $\partial_0$ for the KK modes, and zero when it acts on a zero mode. What we really should do, is to separate this Lagrangian into a zero mode part (5d SYM) and a KK tower, but we may condense the notation if we define $\partial_0^{-1}$ to be zero when it acts on 5d SYM fields. 

If we vary $A_m$ in this Lagrangian, then we get the Maxwell equation of motion, which is a consequence of the selfduality equation. If we vary $B_{mn}$ for the KK modes then we actually get the selfduality equation of motion itself. 

A term like $\bar\psi\gamma^m \psi \partial_m r = 0$ is identically zero. So such a term could never produce the second term $- \frac{1}{2r} (\partial_m r) \gamma^m \psi$ in the fermionic equation of motion. This shows that the prefactor $1/r$ in front of the Lagrangian can be obtained from supersymmetry alone, once $r$ has been made space-dependent.

\section{Mode expansion}
We introduce a discrete basis of real-valued $2\pi$ periodic functions $\varphi^a(t)$ on the circle fiber, where $a\in \mb{Z}$. Since it is a basis, we can expand the derivative in the basis functions as 
\bea
\partial_0 \varphi^a(t) &=& \varphi^b(t) T_b{}^a
\eea
for some matrix $T_b{}^a$. If we define the metric
\bea
h^{ab} &=& \int dt \varphi^a(t)\varphi^b(t)
\eea
then we find the constraints 
\bea
T^{ab} + T^{ba} &=& 0
\eea
where
\bea
T^{ab} &=& T^a{}_c h^{cb}
\eea
We have the completeness relation
\bea
h_{ab} \varphi^a(t') \varphi^b(t) &=& \delta(t-t')
\eea
where $h_{ab}$ denotes the inverse metric. Any function on the circle can be mode expaned as
\bea
f(t) &=& f_a \varphi^a(t)
\eea
where the coefficients are
\bea
f_a &=& \int dt f(t) \varphi_a(t)
\eea
If we have a function on the form
\bea
v(t) &=& v_{ab...} \varphi^a(t) \varphi^b(t) ...
\eea
then we get
\bea
\partial_0 v(t) &=& T[v_{ab...}] \varphi^a(t) \varphi^b(t) ...
\eea
where we define
\bea
T[v_{ab...}] &=& T_a{}^{a'}v_{a'b...}+T_b{}^{b'}v_{ab'...} + ...
\eea
In particular, we find that the metric is time translation invariant,
\bea
T[h_{ab}] &=& 0
\eea
We define 
\bea
f_{abc...d} &=& \int dt \varphi_a(t) \varphi_b(t) \varphi_c(t)...\varphi_d(t) 
\eea

\section{A nonlocal time derivative operator}
Let us consider a function $f(t)$ of time $t$. We may define the time derivative of this function as a limit
\bea
T[f(t)] &=& \lim_{\eps \rightarrow 0} \frac{f(t+\eps) - f(t)}{\eps}
\eea
We will also use the conventional notation
\bea
T[f(t)] &=& \partial_t f(t)
\eea
The function as well as its time derivative are local, because they are defined locally at each point $t$. We may generalize the concept of a local function of time, to a nonlocal function of time $f(t',t)$ which depends on time through two time points $t'$ and $t$ that may or may not be equal. We would now like to define a time derivative of this nonlocal function. If we do not discriminate among $t$ and $t'$, then the natural generalization from the local to the nonlocal time derivative, will be as the following limit 
\bea
T[f(t',t)] &=& \lim_{\eps \rightarrow 0} \frac{f(t+\eps,t'+\eps) - f(t,t')}{\eps}
\eea
or in other words,
\bea
T[f(t',t)] &=& \partial_{t'} f(t',t) + \partial_t f(t',t)
\eea
We do not need to worry about reparametrization invariance of our definition of $T[f(t',t)]$, because for our circle bundle the time coordinate is fixed and can not be reparametrized without taking us outside the circle bundle form of the metric. For the special case that $f(t',t) = g(t') h(t)$ is a product of two local functions, we get
\bea
T[g(t')h(t)] &=& \partial_{t'} g(t') h(t) + g(t') \partial_t h(t)
\eea
and we see that the limit $t'\rightarrow t$ is smooth, and in the limit we recover the chain rule for the time derivative of a product of two functions, 
\bea
T[g(t)h(t)] &=& \partial_t g(t) h(t) + g(t) \partial_t h(t)
\eea

Let us now consider an infinitesimal supersymmetry variation
\bea
\delta_{\eps(t)} \Phi(t)
\eea
of some field $\Phi(t)$ depending locally on $t$. In the Lagrangian there may appear a time derivative of this field, and therefore we will need to understand what is a local variation of that time derivative as well. We define
\bea
\delta_{\eps(t)} T[\Phi(t)] &=& T[\delta_{\eps(t)} \Phi(t)]
\eea
To make this more concrete, let us assume that the variation depends linearly on $\eps(t)$ as
\bea
\delta_{\eps(t)} \Phi(t) &=& \eps(t) \Psi(t)
\eea
We then have
\ben
\delta_{\eps(t)} T[\Phi(t)] = T[\eps(t) \Phi(t)] = \partial_t \eps(t) \Phi(t) + \eps(t) \partial_t \Phi(t)\label{chainrule}
\een
In the last step we have used the chain rule for the time derivative. 

We would now like to introduce a nonlocal supersymmetry variation
\bea
\delta_{\eps(t')} \Phi(t) &=& \eps(t') \Psi(t)
\eea
The Lagrangian is the same local Lagrangian, and it may contain the local time derivative $T[\Phi(t)] = \partial_t \Phi(t)$ of that field. Now we would like to ask ourselves how this will vary when we act by the nonlocal supersymmtry variation,
\bea
\delta_{\eps(t')} T[\Phi(t)] = ?
\eea
If we define this as follows
\bea
\delta_{\eps(t')} T[\Phi(t)] = \bar\eps(t') \partial_t \Psi(t)
\eea
then we will run into the problem of having a discontinuity as $t'$ approaches $t$. Namely, when $t' = t$, we have by the chain rule the result in (\ref{chainrule}), which means that our equations will be inconsistent, or we will need a different set of equations when $t' \neq t$ as compared to when $t' = t$. Therefore we will instead propose that we shall not view $T$ that appears in the Lagrangian as the local time derivative $\partial_t$, but rather as a more general time derivative operator as we defined it above, and which reduces to the local time derivative when it acts on a local field. Thus our proposal is that for the nonlocal supersymmetry variation, the time derivative of a field will vary according to 
\bea
\delta_{\eps(t')} T[\Phi(t)] = T[\delta_{\eps(t')} \Phi(t)] = \partial_{t'} \eps(t') \Psi(t) + \bar\eps(t') \partial_t \Psi(t)
\eea
This definition is now consistent in the sense that it applies to both the case that $t' \neq t$ as well as to the case when $t' = t$, where the usual chain rule of differention can be applied. 

In terms of modes, this relation reads
\ben
\delta_a T[\Phi_b] = T[\delta_a \Phi_b] = T_a{}^c \delta_c \Phi_b + T_b{}^c \delta_a \Phi_c\label{nl}
\een
where we define
\bea
\delta_a &=& \int dt' \varphi_a(t')\delta_{\eps(t')}
\eea

\section{Abelian nonlocal supersymmetry}
We mode expand the supersymmetry parameter as
\bea
\eps(t) &=& \eps_a \varphi^a(t)
\eea
The Killing spinor equation on the modes becomes
\bea
\t D_m \eps_a &=& \gamma_m \eta_a - \frac{r}{4} w_{mn} \gamma^n\eps_a+\kappa_m T_a{}^b\eps_b\cr
\eta_a &=& \frac{1}{r} T_a{}^b \eps_b - \frac{r}{8} w_{mn}\gamma^{mn} \eps_a + \frac{1}{2r}(\partial_m r) \gamma^m \eps_a
\eea
We derive the identities\footnote{Round brackets means symmetrization and square brackets means antisymmetrization, all with weight one.}
\ben
\partial_m s_{ef} &=& r w_{mn} s^n_{ef} + \frac{1}{r} (\partial_m r) s_{ef} \cr
&&- \frac{1}{r} T[s_{m,ef}] + \kappa_m T[s_{ef}]\label{id1}\\
\partial_m (r s_{ef}) + 4 r \bar\eps_e \gamma_m \eta_f &=& T[s_{m,ef}] + r \kappa_m T[s_{ef}]\label{id2}\\
\partial_{[m} s_{n],ef} + 2 \bar\eps_e \gamma_{mn} \eta_f &=& \frac{r}{2} s_{ef} w_{mn} + \kappa_{[m} T[s_{n],ef}]\label{id3}
\een
Here we define
\bea
s_{ef} &=& - \bar\eps_{(e}\eps_{f)}\cr
s^m_{ef} &=& \bar\eps_{(e}\gamma^m\eps_{f)}
\eea

The nonlocal abelian supersymmetry variations read
\bea
\delta_a \sigma_b &=& - i \bar\eps_a \sigma_b\cr
\delta_a A_{m,b} &=& i r \bar\eps_a \gamma_m \chi_b\cr
\delta_a B_{mn,b} &=& i \bar\eps_a \gamma_{mn} \chi_b - 2 i r \kappa_{[m} \bar\eps_a \gamma_n \chi_b \cr
\delta_a \chi_b &=& \frac{1}{2r} \gamma^{mn} \eps_a \F_{mn,b} + \gamma^m \eps_a  \(D_m\sigma_b - T_b{}^c \kappa_m \sigma_c\) + \frac{1}{r} T_b{}^c \eps_a\sigma_c + 4\eta_a\sigma_b 
\eea
The abelian action (\ref{abelL}) is invariant under these nonlocal variations, and hence they are symmetry variations. But for this we must interpret the local time derivatives that appears in the action as time derivative operators with the property (\ref{nl}).

The closure relations are
\bea
\frac{1}{2}\{\delta_f,\delta_e\} \sigma_b &=& - i \L_{S,ef} \sigma_{b} - 4 i \bar\eps_e\eta_f \sigma_b\cr
\frac{1}{2}\{\delta_f,\delta_e\} A_{m,b} &=& - i \L_{S,ef} A_{m,b} + \(\partial_m \Lambda_{0,efb} - \kappa_m T[\Lambda_{0,efb}]\) - T[\t\Lambda_{m,efb}]\cr
\frac{1}{2}\{\delta_f,\delta_e\} B_{mn,b} &=& - i \L_{S,ef} B_{mn,b} + 2 \partial_m \(\t\Lambda_{n,efb} + \kappa_m \Lambda_{0,efb}\)\cr
&& + i s^r_{ef} \E[B_b]_{rmn}\cr
\frac{1}{2}\{\delta_f,\delta_e\}\chi_b &=& - i \L_{S,ef} \chi_b - 5 i \bar\eps_e\eta_f \chi_b + \frac{i r}{8} s^{mn,ij}_{ef} w_{mn} \tau^{ij} \chi_b\cr
&& + \(- \frac{3 i}{8} s_{ef}+ \frac{3 i}{8} s_{p,ef} \gamma^p + \frac{i}{64} s^{ij}_{pq,ef} \gamma^{pq} \tau^{ij}\)\E[\chi_b]
\eea
where
\bea
\Lambda_{0,efb} &=& i \(s^m_{ef} A_{m,b} - r s_{ef} \sigma_b\)\cr
\t\Lambda_{m,efb} &=& \frac{i}{r} \(s_{ef} A_{m,b} - r s_{m,ef} \sigma_b\) + i s^n_{ef} \(\kappa_n A_{m,b} - \kappa_m A_{n,b}\) - i B_{mn,b} s^n_{ef}
\eea
Here $\L_{S,ef} = \L_{S_{ef}}$ denotes the 6d Lie derivative along the 6d vector field $S^M_{ef} = \bar\eps_e \Gamma^M \eps_f$. We can extract the 6d conformal Killing vector field $S^M(t) = \bar\eps(t) \Gamma^M \eps(t)$ by contracting with $\varphi^e(t) \varphi^f(t)$. Still the resulting conformal transformation, the Lie derivative acting on the field, will be nonlocal because the field has to be in general evaluated at a different time from the time $t$. Thus closure results in conformal transformations of the nonlocal type $\L_{S(t)} \Phi(t')$. This shall be contrasted with local variations, which result in closure relations on the form
\bea
\delta^2 \Phi_a &=& f_a{}^{efg} \L_{S,ef} \Phi_g
\eea
By contracting this relation with $\varphi^a(t)$ and noting the identity
\bea
\varphi^a(t) f_a{}^{efg} &=& \varphi^e(t) \varphi^f(t) \varphi^g(t)
\eea
we get
\bea
\delta^2 \Phi(t) &=& \L_{S(t)} \Phi(t)
\eea
where $S(t) = \bar\eps(t) \Gamma^M \eps(t)$. This is the usual local conformal transformation.

\section{Nonabelian nonlocal supersymmetry}
Having the set the ground, we are now ready to present an ansatz for the nonabelian generalization of the nonlocal superconformal variations. For the zero modes we take the nonlocal variations to be
\bea
\delta_a \phi &=& - i \bar\eps_a \chi\cr
\delta_a a_m &=& i r \bar\eps_a \gamma_m \psi\cr
\delta_a \psi &=& \frac{1}{2r} \gamma^{mn} \eps_a f_{mn} + \gamma^m \eps_a D_m \phi + 4 \eta_a \phi 
\eea
As we advertised in the introduction, we see that with a nonlocal variation we are able to express these variations in a close form such that the variation only involves the zero mode fields. For a local variation we are instead forced to pick the zero mode $\eps_0$ (for which $\partial_0 \eps_0 = 0$) as the supersymmety parameter and thereby we reduce the  symmetry to $\C_{5d,(1,0)}$. It is only by allowing for the variation to be nonlocal that we can allow the supersymmetry parameter to carry a nonvanishing mode number $\eps_a$ whose time derivative $T_a{}^b \eps_b$ is nonvanishing and then we can realize $\C_{6d,(1,0)}$. 

For the KK modes, we make the ansatz
\bea
\delta_a \sigma_b &=& - i \bar\eps_a \chi_b\cr
\delta_a A_{m,b} &=& i r \bar\eps_a \gamma_m \chi_b\cr
\delta_a B_{mn,b} &=& i \bar\eps_a \gamma_{mn} \chi_b - 2 i r \kappa_{[m} \bar\eps_a \gamma_n \chi_b  + C_b{}^c \([\phi,\bar\eps_a\gamma_{mn}\chi_c]-[\sigma_c,\bar\eps_a\gamma_{mn}\psi]\) + \t C_b{}^c [A_{n,c},\bar\eps_a\gamma_m\psi]\cr
\delta_a \chi_b &=& \frac{1}{2r} \gamma^{mn} \eps_a \F_{mn,b} + \gamma^m \eps_a  \(D_m\sigma_b - T_b{}^c \kappa_m \sigma_c\) + \frac{1}{r} T_b{}^c \eps_a\sigma_c + 4\eta_a\sigma_b - i r \eps_a [\phi,\sigma_b]
\eea
Let here illustrate how we determine the coefficients $C_b{}^c$ and $\t C_b{}^c$. We then begin by considering 
\bea
\delta_f \delta_e B_{mn,b} &=& - 2 i r \kappa_m \bar\eps_e \gamma_n \delta_f \chi_b + C_b{}^c [\phi,\bar\eps_e \gamma_{mn} \delta_f \chi_c] + ...
\eea
where we insert the variation
\bea
\delta_f \chi_b &=& \gamma^m \eps_f T_b{}^c \kappa_m \sigma_c - i r \eps_f [\phi,\sigma_b] + ...
\eea
We then get
\bea
\delta_f \delta_e B_{mn,b} &=& - 2 r^2 \kappa_m \bar\eps_e \gamma_n \eps_f [\phi,\sigma_b] - 2 C_b{}^c T_c{}^d \kappa_m \bar\eps_e \gamma_n \eps_f [\phi,\sigma_d] + ...
\eea
For these terms to cancel, we shall take
\bea
C_a{}^b &=& - r^2 (T^{-1})_a{}^b
\eea
Let us now look at a term
\bea
\delta_f \delta_e B_{mn,b} &=& - i \bar\eps_e \gamma^r \eps_f H_{rmn,b} + ...
\eea
that arises when we use a selfduality equation of motion. Here we define
\bea
H_{rmn,b} &=& 3 D_r B_{mn} - i (T^{-1})_b{}^c [A_{r,c},f_{mn}] - 2 i (T^{-1})_b{}^c [A_{m,c},f_{nr}]
\eea
The last term gives rise to a term 
\bea
\delta_f \delta_e B_{mn,b} &=& - 2 \bar\eps_e \gamma^r \eps_f (T^{-1})_b{}^c [A_{m,c},f_{nr}] + ...
\eea
that we need to cancel. We cancel it by varying the term in $\delta_e B_{mn,b}$ that is proportional to $\t C_b{}^c$, which will produce a term of the form
\bea
\delta_f \delta_e B_{mn,b} &=& - \frac{1}{r} \bar\eps_e \gamma^r \eps_f \t C_b{}^c [A_{m,c},f_{nr}] + ...
\eea
We see that for the two terms to cancel, we shall take
\bea
\t C_a{}^b &=& - 2 r (T^{-1})_a{}^b
\eea
We have now found that we should have 
\bea
\delta_a B_{mn,b} &=& i \bar\eps_a \gamma_{mn} \chi_b - 2 i r \kappa_{m} \bar\eps_a \gamma_n \chi_b\cr
&& + r^2 (T^{-1})_b{}^c \([\sigma_c,\bar\eps_a\gamma_{mn}\psi]-[\phi,\bar\eps_a\gamma_{mn}\chi_c]\) - 2 r (T^{-1})_b{}^c [A_{n,c},\bar\eps_a\gamma_m\psi]
\eea
and then the closure relations become
\bea
\frac{1}{2}\{\delta_f,\delta_e\} \sigma_b &=& - i \L_{S,ef} \sigma_b - 4 i \bar\eps_e \eta_f \sigma_b - i [\sigma_b,\lambda_{ef}]\cr
\frac{1}{2}\{\delta_f,\delta_e\} A_{m,b} &=& - i \L_{S,ef} A_{m,b} + \(D_m \Lambda_{0,efb} - \kappa_m T[\Lambda_{0,efb}]\) - T[\t\Lambda_{m,efb}] - i [A_{m,b},\lambda_{ef}]\cr
\frac{1}{2}\{\delta_f,\delta_e\} B_{mn,b} &=& - i \L_{S,ef} B_{mn,b}\cr
&& + 2 D_{[m} \(\t\Lambda_{n],efb}+\kappa_n\Lambda_{0,efb}\) - i [B_{mn,b},\lambda_{ef}] + i [f_{mn},T^{-1}[\Lambda_{0,efb}]]\cr
&& + i s^r \E[B]_{rmn,b}\cr
\frac{1}{2}\{\delta_f,\delta_e\} \chi_b &=& - i \L_{S,ef} \chi_b - 5 i \bar\eps_e \eta_f \chi_b + \frac{i r}{8} s^{mn,ij}_{ef} w_{mn} \tau^{ij} \chi_b\cr
&& - i[\chi_b,\lambda_{ef}]\cr
&& + \(- \frac{3 i}{8} s_{ef} + \frac{3 i}{8} s_{p,ef} \gamma^p + \frac{i}{64} s^{ij}_{pq,ef} \gamma^{pq} \tau^{ij}\)\E[\chi_b]\label{FEOM}
\eea
where
\bea
\E[B]_{rmn,b} &=& H_{rmn,b} \cr
&&+ \frac{1}{2r} \E_{rmn}{}^{pq} \(\F_{pq}-ir^2 (T^{-1})_b{}^c([\sigma_c,f_{pq}] - [\phi,\F_{pq,c}]) + \frac{r^3}{2}(T^{-1})_b{}^c\{\bar\psi\gamma_{pq}\chi_c\}\) \cr
&& - 3 \kappa_r \F_{mn,b}\cr
\E[\chi_b] &=& \gamma^m (\t D_m \chi_b - \kappa_m T_b{}^c \chi_c) - \frac{1}{r} T_b{}^c \chi_c + \frac{1}{2r} (\partial_m r) \gamma^m \chi_b - \frac{r}{8} w_{mn} \gamma^{mn} \chi_b\cr
&& - i r [\chi_b,\phi] + 2 i r[\psi,\sigma_b]
\eea
and
\bea
\lambda_{ef} &=& i \(s^q_{ef} a_q - r s_{ef}\phi\)\cr
\Lambda_{0,efb} &=& i \(s^q_{ef} A_{q,b} - r s_{ef} \sigma_b\)\cr
\t\Lambda_{m,efb} &=& - i B_{mn,b} s^n_{ef} - i \(s_{m,ef} \sigma_b - \frac{s_{ef}}{r} A_{m,b}\)\cr
&& + i s^n_{ef} \(\kappa_n A_{m,b} - \kappa_m A_{n,b}\) - r T^{-1}\[\phi,r s_{m,ef}\sigma_b - s_{ef} A_{m,b}\]
\eea

\section{Discussion}
One application that our result might have is to resolve the conflict between the two proposals \cite{Douglas:2010iu}, \cite{Lambert:2010iw} and \cite{Ho:2011ni}, \cite{Bonetti:2012st}, \cite{Gustavsson:2018pov}. The conflict may get resolved by utilizing the nonlocal superconformal transformation. We have seen that it seems impossible to realize the $\C_{6d,(1,0)}$ symmetry locally in a classical field theory. There may be other ways this symmetry can get realized. One possiblilty is that $\C_{6d,(1,0)}$ emerges at quantum level by means of instanton particles and enhance the classical symmetry $\C_{5d,(1,0)}$ that was shown to be present at the classical level in 5d SYM in \cite{Gustavsson:2018rcc}. This would be similar to how $\N=8$ emerges at quantum level by means of monopole operators from $\N=6$ ABJM theory. What plays the role of ABJM for M2 would then be 5d SYM for M5. Another possibility, which need not be in conflict with this instanton-particle proposal, is if there is a classical field theory in which $\C_{6d,(1,0)}$ is realized in a nonlocal way. If we demand an ordinary local field theory description of the 6d (1,0) tensor multiplet, then the only possibiltiy seems to be a 5d SYM, or something similar (for example lightcone reduction of 6d theory), that has a subgroup of $\C_{6d,(1,0)}$ realized at a classical level. This does not rule out the possibilty that there may exist ways to realize $\C_{6d,(1,0)}$ classically, only that we may not be able to realized it as a local symmety in a local field theory. 

Regarding the proposal in \cite{Ho:2011ni}, \cite{Bonetti:2012st}, \cite{Gustavsson:2018pov}, we would now like to argue that this proposal corresponds to a nonlocal field theory. This can be seen from three places. Gauge symmetry is nonlocal, the selfduality equation of motion is nonlocal. But even if that is not convincing enough (one may perhaps object by saying that neither $H_{rmn}$ nor the gauge symmetry are really needed in the description of the theory?), then we have now provided a third indication that the theory is nonlocal. Namely even if the theory is an ordinary local field theory on $S^6$, it would necessarily become nonlocal if we apply a nonlocally realized conformal transformation to map this theory to $\mb{R}^6$. Then we will get a nonlocal theory on $\mb{R}^6$ because the transformation is nonlocal, and in particular that means that 6d Lorentz symmery will be realized in a nonlocal way on $\mb{R}^6$. 

Let us discuss the question of how to obtain $(2,0)$ supersymmetry. We have failed with a manifestly $SO(5)$ covariant ansatz for the supersymmetry with a KK tower (unless of course the gauge group is abelian). We then need a unit vector $v^A$ that selects a direction on $S^4$ and breaks $(2,0)$ supersymmetry down to $(1,0)$. We may hopefully be able to extract a nonabelian action $S[v^A,\phi^A]$ from our results (this would be rather straightforward) from which we may get an $SO(5)$ invariant action by integrating over $v^A \in S^4$,
\bea
\int_{S^4} dv e^{-S[v,\phi]} &=& e^{-S_{eff}[\phi]}
\eea
Then the question is if $S_{eff}$ will also be $(2,0)$ supersymmetric. This seems to be a difficult question. 

Another interesting application of the circle-bundle formulation of the 6d tensor multiplet is to singular fibrations \cite{Witten:2009at}, \cite{Ohlsson:2012yn}, \cite{Lambert:2018mfb}. The circle fiber may shrinks to zero size at a submanifold of dimensions $0$, $2$ and $4$ respectively \cite{Witten:2009at}. We can see those dimensions appear from the following sequence of conformally flat six-manifolds, $S^6$, $H^2\times S^4$ and $H^4 \times S^2$. Odd dimensional spheres have the Hopf fibration and are regular fibrations. But even dimensional spheres necessarily have singular fibrations. It will be interesting to derive the corresponding theories that we may need to supplement to 5d SYM and that live at those singular points or submanifolds. Perhaps one may use results from \cite{Belyaev:2008xk} that provides a machinery to derive supersymmetric boundary theories. Here we have singular loci, and they are not quite boundaries though.

\subsection{Acknowledgments}
I would like to thank Yang Zhou, Jeong-Hyuck Park and Pei-Ming Ho for discussions. This work was supported in part by NRF Grant 2017R1A2B4003095.

\appendix

\section{Closure on $\sigma_b$} 
We get
\bea
\frac{1}{2}\{\delta_f,\delta_e\} \sigma_b &=& - i \bar\eps_e\gamma^m\eps_f \(D_m \sigma_b - T_b{}^c \kappa_m \sigma_c\) - \frac{i}{r} \bar\eps_e \eps_f T_b{}^c \sigma_c \cr
&& - 4 i \bar\eps_e \eta_f \sigma_b\cr
&& - r \bar\eps_e \eps_f [\phi,\sigma_b]  
\eea
We will write this as
\bea
\frac{1}{2}\{\delta_f,\delta_e\} \sigma_b &=& - i \L_{S,ef} \sigma_b - 4 i \bar\eps_e \eta_f \sigma_b - i [\sigma_b,\lambda_{ef}]
\eea
where
\bea
\lambda_{ef} &=& i \(s^m_{ef} a_m - r s_{ef} \phi\)
\eea

\section{Closure on $A_{m,b}$}
We divide the computation into three pieces, 
\bea
\delta_f \delta_e A_{m,b} &=& i r \bar\eps_e \gamma_m \(\delta_f \chi_b + \delta'_f \chi_b + \delta''_f \chi_b\)
\eea
where
\bea
\delta_a \chi_b &=& \frac{1}{2r} \gamma^{mn} \eps_a \F_{mn,b}\cr
\delta'_a \chi_b &=& \gamma^m \eps_a  \(D_m\sigma_b - T_b{}^c \kappa_m \sigma_c\) + \frac{1}{r} T_b{}^c \eps_a\sigma_c + 4\eta_a\sigma_b\cr
\delta''_a \chi_b &=& - i r \eps_a [\phi,\sigma_b]
\eea
We get
\bea
\delta_f \delta_e A_{m,b} &=& i s^q_{ef} \F_{mq,b}
\eea
where we define
\bea
\F_{mn,a} &=& 2D_m A_{n,a} + T_a{}^b B_{mn,b}
\eea
Then
\bea
\delta_f \delta_e A_{m,b} &=& - i \L_{S,ef} A_{m,b} + D_m\(i s^q_{ef} A_{q,b}\)\cr
&& - T[\frac{i}{r} s_{ef} A_{m,b} + i s^q_{ef} \kappa_q A_{m,b} + i s^q_{ef} B_{qm}]\cr
&& + [A_{m,b},s^q_{ef}]
\eea
Next
\bea
\delta'_f \delta_e A_{m,b} &=& - i r s_{ef} \(D_m \sigma_b - T_b{}^c \kappa_m \sigma_c\) + i s_{m,ef} T_b{}^c \sigma_c + 4 i r \bar\eps_e \gamma_m \eta_f \sigma_b
\eea
We extract a total derivative,
\bea
\delta'_f \delta_e A_{m,b} &=& D_m \(- r s_{ef} \sigma_b\) + i r s_{ef} T_b{}^c \sigma_c \kappa_m + i s_{m,ef} T_b{}^c \sigma_c\cr
&& + i \partial_m (r s_{ef}) + 4 i r \bar\eps_e \gamma_m \eta_f \sigma_b
\eea
Now we apply (\ref{id2}) on the second line and get
\bea
\delta'_f \delta_e A_{m,b} &=& D_m \(- r s_{ef} \sigma_b\) - \kappa_m \(-i r  T[s_{ef}\sigma_b]\) - T[- i s_{m,ef}\sigma_b]
\eea
Let us finally extract all commutators. One is from $\delta_f\chi_b$ and another is from $\delta''_f\chi_b$,
\bea
(\delta_f\delta_e A_{m,b})_{comm} &=& [A_{m,b},s^n_{ef}a_n] + r^2 s_{m,ef}[\phi,\sigma_b]\cr
&=& [A_{m,b},s^n_{ef} a_n - r s_{ef} \phi] - r [\phi,s_{ef} A_{m,b} - r s_{m,ef}\sigma_b]\cr
&=& - i [A_{m,b},\lambda_{ef}] - T[\t\Lambda_{m,efb}] 
\eea
where
\bea
\t\Lambda_{m,efb} &=& r T^{-1}[\phi,s_{ef} A_{m,b} - r s_{m,ef}\sigma_b]
\eea

\section{Closure on $A_m$, local computation}
We here present part of the closure computation for local $\C_{5d,(1,0)}$ superconformal symmetry variations where $\partial_0 \eps = 0$ and the variations are
\bea
\delta \sigma &=& - i \bar\eps \chi\cr
\delta A_m &=& i r \bar\eps \gamma_m \chi\cr
\delta B_{mn} &=& i \bar\eps \gamma_{mn} \chi - 2 i r \kappa_{[m} \bar\eps\gamma_n \chi  + \frac{i r^2}{n} \([\sigma,\bar\eps\gamma_{mn}\psi]-[\phi,\bar\eps\gamma_{mn}\chi]\) - \frac{2 i r}{n} [A_n,\bar\eps\gamma_m\psi]\cr
\delta \chi &=& \frac{1}{2r} \gamma^{mn} \eps \F_{mn} + \gamma^m \eps  \(D_m\sigma - in\kappa_m \sigma\) + \frac{in}{r}\eps\sigma + 4\eta\sigma - i r \eps [\phi,\sigma]
\eea
This way we can see that the corresponding nonlocal computation is almost completely analogous, and that $i n$ gets replaced by $T$. We get
\bea
\delta^2 A_m &=& i s^n \F_{mn} - i r s \(D_m \sigma - i n \kappa_m \sigma\) - n s_m \sigma + 4 i r \bar\eps\gamma_m\eta\sigma + r^2 \bar\eps\gamma_m\eps s_m [\phi,\sigma]
\eea
We rewrite this as
\bea
\delta^2 A_m &=& - i \L_S A_m + D_m \(i(s^q A_q - r s \sigma)\) \cr
&& + n s^n \kappa_n A_m - n \kappa_m r s \sigma\cr
&& - n \(B_{mn}s^n + s_m\sigma - \frac{s}{r} A_m - \frac{1}{n}r^2 s_m [\phi,\sigma]\) + [A_m,s^q a_q]\cr
&& + i \partial_m(r s)\sigma + 4 i r \bar\eps\gamma_m\eta \sigma
\eea
where we have extracted the 6d Lie derivative. We note that the last line is zero as a consequence of the 5d Killing spinor equation. We add and subtract $[A_m,rs\phi]$ and get
\bea
\delta^2 A_m &=& - i \L_S A_m + D_m \(i(s^q A_q - r s \sigma)\)\cr
&& + n s^n \kappa_n A_m - n r s \kappa_m \sigma \cr
&& - n \(B_{mn}s^n + s_m\sigma - \frac{s}{r}A_m + \frac{1}{n}r [\phi,s A_m - r s_m\sigma]\) + [A_m,s^q a_q-rs\phi]
\eea
We can write this in the form
\bea
\delta^2 A_m &=& - i \L_S A_m + D_m \Lambda_0 - i n \kappa_m \Lambda_0 - i n \t\Lambda_m - i [A_m,\lambda]
\eea
where 
\ben
\lambda &=& i (s^q a_q - r s \phi)\cr
\Lambda_0 &=& i (s^q A_q - r s \sigma)\cr
\t\Lambda_m &=& - i s^n \(\kappa_m A_n - \kappa_n A_m\) - i B_{mn} s^n + \frac{i}{r} \(s A_m - r s_m \sigma\) \cr
&& - \frac{i r}{n} [\phi,s A_m - r s_m \sigma]\label{tilde}
\een

\section{Closure on $B_{mn}$, local computation}
We get
\bea
\delta^2 B_{mn} &=& \frac{is}{r} \F_{mn} + \frac{i}{2r} s^r \E_{mnr}{}^{pq} \F_{pq} - 2 i \kappa_m s^q \F_{nq}\cr
&&+ 2 D_m \(-i(s_n+rs\kappa_n)\sigma\)\cr
&& \underbrace{- 2 r^2 \kappa_m s_n \eps [\phi,\sigma]}_W + \underbrace{\frac{ir^2}{n}[\sigma,\bar\eps\gamma_{mn}\delta \psi]}_Y \underbrace{- \frac{ir^2}{n}[\phi,\bar\eps\gamma_{mn}\delta\chi]}_Z \underbrace{- \frac{2ir}{n}[A_n,\bar\eps\gamma_m\delta\psi]}_X
\eea
The contributions that appear in the first and the third lines are obvious, but to understand the contribution on the second line require some computation. We get this result from
\bea
(\delta^2 B_{mn})_{2nd} &=& i \bar\eps\gamma_{mn} \delta\chi - 2 i r \kappa_m \bar\eps \gamma_n \delta\chi
\eea
where we plug in $\delta \chi = \gamma^p \eps (D_p \sigma - \kappa_p \partial_0 \sigma) + 4 \eta \sigma + ... $ and omit the dots. Then we get
\bea
(\delta^2 B_{mn})_{2nd} &=& 2 D_m(-i s \sigma) + 2 i (\partial_m s_n) \sigma + 4 i \bar\eps\gamma_{mn}\eta\sigma \cr
&& + 2 i r \kappa_m (s D_n \sigma - 4 \bar\eps\gamma_n\eta\sigma)
\eea
Now we use (\ref{id2}) and (\ref{id3}) to get after some computation the result
\bea
(\delta^2 B_{mn})_{2nd} &=& 2 D_m (- i s_n \sigma - i r \kappa_n s \sigma)
\eea
which is the second line. 

Let us now consider the first line,
\bea
\delta^2 B_{mn} &=& \frac{i}{r} s \F_{mn} + \frac{i}{2r} s^r \E_{rmn}{}^{pq} \F_{pq} - 2 i \kappa_m s^r \F_{nr}
\eea
We assume some selfduality relation
\bea
\frac{1}{2r} \E_{rmn}{}^{pq} \F_{pq} &=& - H_{rmn} + 3 \kappa_r \F_{mn} + X_{rmn}
\eea
where $X_{rmn}$ will be determined from requiring on-shell closure. We then get
\bea
\delta^2 B_{mn} &=& \frac{is}{r} \F_{mn} - i s^r H_{rmn} + i s^r \kappa_r \F_{mn} + i s^r X_{rmn}
\eea
Then we use the definition
\bea
H_{rmn} &=& 3 D_r B_{mn} - \frac{3}{n} [f_{mn},A_r]
\eea
and we get
\bea
\delta^2 B_{mn} &=& - i \L_S B_{mn} + 2 D_m \Lambda_n + [s^r a_r,B_{mn}]+ \frac{i s^r}{n} [f_{mn},A_r] + \frac{2 i s^r}{n} [f_{rm},A_n] + i s^r X_{rmn}
\eea
We can write this in the form
\bea
\delta^2 B_{mn} &=& - i \L_S B_{mn} + 2 D_m \Lambda_n - i [B_{mn},\lambda] + \frac{1}{n} [f_{mn},\Lambda_0]\cr
&& + \frac{2 i s^r}{n} [f_{rm},A_n] + i s^r X_{rmn}
\eea
where
\bea
\Lambda'_m &=& \frac{i}{r} \(sA_m - rs_m\sigma\) - i s^r B_{mr} + i s^r \kappa_r A_m-ir\kappa_m s \sigma\cr
\Lambda'_0 &=& is^r A_r\cr
\lambda' &=& i s^r a_r
\eea
We now notice that 
\bea
\Lambda'_m &=& \t\Lambda_m + \kappa_m \Lambda_0
\eea
where $\t\Lambda_m$ and $\Lambda_0$ are given in (\ref{tilde}).

Let us now study the term
\bea
X &:=& - \frac{2 i r}{n} [A_n,\bar\eps\gamma_m\delta\psi]
\eea
where
\bea
\delta \psi &=& \frac{1}{2r} \gamma^{pq} \eps f_{pq}+\gamma^p\eps D_p\phi + 4 \eta \phi
\eea
We get
\bea
X &=& - \frac{2 i}{n} [A_n,s^q f_{mq}] - \frac{2 i}{n} [A_n,r\bar\eps\eps D_m\phi] - \frac{8 i r}{n} [A_n,\bar\eps\gamma_m\eta \phi]
\eea
By using the 5d Killing spinor equation this can be recast in the form
\bea
X &=& - \frac{2 i}{n} [A_n,s^q f_{mq}] - \frac{2 i}{n} [A_n,D_m(r \bar\eps \eps \phi)]
\eea
We have two more terms
\bea
Y = \frac{ir^2}{n}[\sigma,\bar\eps\gamma_{mn}\delta \psi] &=& \frac{i r s}{n} [\sigma,f_{mn}] + \frac{i r s^r}{2n}\E_{rmn}{}^{pq}[\sigma,f_{pq}]\cr
&& + \frac{2 i r^2}{n} s_m [\sigma,D_n\phi] + \frac{4 i r^2}{n}\bar\eps\gamma_{mn}\eta [\sigma,\phi]
\eea
and
\bea
Z = -\frac{ir^2}{n}[\phi,\bar\eps\gamma_{mn}\delta\chi] &=& -\frac{irs}{n}[\phi,\F_{mn}] - \frac{irs^r}{2n}\E_{rmn}{}^{pq}[\phi,\F_{pq}]\cr
&& - \frac{2 i r^2}{n} s_m [\phi,D_n\sigma] - 2 r^2 s_m\kappa_n[\phi,\sigma] - \frac{4 i r^2}{n}\bar\eps\gamma_{mn}\eta[\phi,\sigma]
\eea
Let us first notice that the fourth term in $Z$ beautifully cancels against the term we called $W$ above. Next we collect the two terms
\bea
\frac{2ir^2}{n}s_m([\sigma,D_n\phi]-[\phi,D_n\sigma]) &=& \frac{2ir^2}{n}s_m D_n([\sigma,\phi])
\eea
We also note that two terms add up to give
\bea
\frac{8 i r^2}{n} \bar\eps\gamma_{mn}\eta[\sigma,\phi]
\eea
and finally by expanding out the first term in $Z$ and adding to it the last term in $X$, we get 
\bea
D_m\(-\frac{2irs}{n}[\phi,A_n]\) + [r s \phi,B_{mn}]
\eea
We can now summarize what we have got so far as
\bea
X+Y+Z+W &=& \frac{i r s}{n} [\sigma,f_{mn}] + \frac{i r s^r}{2n}\E_{rmn}{}^{pq}[\sigma,f_{pq}]\cr
&& + D_m\(-\frac{2irs}{n}[\phi,A_n]\) + [r s \phi,B_{mn}] - \frac{irs^r}{2n}\E_{rmn}{}^{pq}[\phi,\F_{pq}]\cr
&& + \frac{2ir^2}{n}s_m D_n([\sigma,\phi]) + \frac{8 i r^2}{n} \bar\eps\gamma_{mn}\eta[\sigma,\phi]\cr
&& - \frac{2 i}{n} [A_n,s^q f_{mq}]
\eea 
By using the 5d Killing spinor equation, we can write the third line above as
\bea
\frac{2ir^2}{n}s_m D_n([\sigma,\phi]) + \frac{8 i r^2}{n} \bar\eps\gamma_{mn}\eta[\sigma,\phi] &=& D_n \(\frac{2 i r^2}{n} s_m [\sigma,\phi]\)
\eea
so that we get
\bea
X+Y+X+W &=& \frac{i r s}{n} [\sigma,f_{mn}] + [r s \phi,B_{mn}] \cr
&& + \frac{i r s^r}{2n}\E_{rmn}{}^{pq}\([\sigma,f_{pq}] - [\phi,\F_{pq}]\)\cr
&& + D_m\(-\frac{2irs}{n}[\phi,A_n]-\frac{2 i r^2}{n} s_n [\sigma,\phi]\)\cr
&& - \frac{2 i}{n} [A_n,s^q f_{mq}]
\eea 
We shall choose 
\bea
X_{rmn} &=& \frac{r}{2n}\E_{rmn}{}^{pq}\([\sigma,f_{pq}] - [\phi,\F_{pq}]\)
\eea
and then the selfduality constraint reads
\bea
H_{rmn} &=& - \frac{1}{2r} \E_{rmn}{}^{pq} \(\F_{pq}-\frac{r^2}{n}([\sigma,f_{pq}] - [\phi,\F_{pq}])\)  + 3 \kappa_r \F_{mn} 
\eea
We now should also obtain the fermionic contribution to this selfdual equation. But that will be the same as in flat space since the variation of those fields is the same as for flat space case. That is so for $\delta A_m$ whose supersymmetry variation does not contain any curvature correction proportional to say $\kappa_m$. We can therefore borrow the result for the fermionic contribution directly from \cite{Gustavsson:2018pov} that was obtained in flat space. 

By adding the various contributions, we obtain the result presented in the main text.

\section{Relations between 6d and 5d quantities}
We have the following 6d gamma matrices as expressed in terms of 5d gamma matrices (decorated with tilde)
\bea
\Gamma_m &=& \t\Gamma_m - r \kappa_m \Gamma^{\h 0}\cr
\Gamma_0 &=& - r \Gamma^{\h 0}\cr
\Gamma^m &=& \t\Gamma^m\cr
\Gamma^0 &=& \frac{1}{r}\Gamma^{\h 0} - \kappa_m \t\Gamma^m\cr
\Gamma^{m0} &=& \frac{1}{r} \t\Gamma^m \Gamma^{\h 0} - \kappa_n \t\Gamma^{mn}\cr
\Gamma_{m0} &=& -r\t\Gamma_m \Gamma^{\h 0}
\eea
We represent the 11d gammas as
\bea
\Gamma^{\h 0} &=& i \sigma^2 \otimes 1 \otimes 1\cr
\Gamma^{\h m} &=& \sigma^1 \otimes \gamma^m \otimes 1\cr
\Gamma^A &=& \sigma^3 \otimes 1 \otimes \tau^A
\eea
and the 6d chirality matrix as
\bea
\Gamma &=& \sigma^3 \otimes 1 \otimes 1
\eea
The 6d covariant derivatives are related to 5d covariant derivatives as follows,
\bea
D_0 \chi &=& i n \chi - \frac{r^2}{8} w_{mn} \gamma^{mn} \chi - \frac{1}{2} (\partial_m r) \gamma^m \chi\cr
D_m \chi &=& \t D_m \chi - i n \kappa_m \chi + \kappa_m D_0 \chi - \frac{r}{4} w_{mn} \gamma^n \chi
\eea
For the supercharges, of opposite 6d chirality, there are some sign changes,
\bea
D_0 \eps &=& i n \eps - \frac{r^2}{8} w_{mn} \gamma^{mn} \eps + \frac{1}{2} (\partial_m r) \gamma^m \eps\cr
D_m \eps &=& \t D_m \eps - i n \kappa_m \eps + \kappa_m D_0 \eps + \frac{r}{4} w_{mn} \gamma^n \eps
\eea
In 5d we shall have 
\bea
(\gamma^{m})^T &=& s C \gamma^m C^{-1}
\eea
with $s = + 1$. This can be understood as follows. For either sign $s = \pm 1$, we get
\bea
(\gamma^5)^T &=& (\gamma^{1234})^T\cr
&=& C \gamma^{1234} C^{-1}\cr
&=& C \gamma^5 C^{-1}
\eea
and so by $SO(5)$ covariance, this should be true for all $\gamma^m$ $(m=1,2,3,4,5)$. 

For 11d gamma matrices we have on the other hand 
\bea
(\Gamma^M)^T &=& - C_{11d} \Gamma^M C_{11d}^{-1}
\eea
and this is not in conflict with the fact that 
\bea
\Gamma^{10} &=& \Gamma^{0123456789}
\eea
since here we find $5$ exhanges and $(-1)^5 = -1$, so any sign $s$ is fine by that argument. An explicit realization is 
\bea
C_{11d} &=& \epsilon \otimes C \times C'
\eea
where we break $SO(1,10) \rightarrow SO(1,5)\times SO(5)$. 

An explicit realization of the $SO(5)_R$ gammas is 
\bea
\tau^I &=& \sigma^1 \otimes \sigma^I\cr
\tau^4 &=& \sigma^2 \otimes 1\cr
\tau^5 &=& -\sigma^3 \otimes 1
\eea
such that we have $\gamma^5 = \gamma^{1234}$. Then we have
\bea
C' &=& \sigma^3 \otimes \epsilon
\eea
and we realize 
\bea
(\gamma^A)^T &=& C' \gamma^A C'^{-1}
\eea

It is important to note that when we use 5d language, we completely dispose of the chirality matrix, and we use 5d spinors which are not Weyl. Then we define in 5d language
\bea
\bar\eps &=& \eps^T C \otimes C'
\eea
with $\eps$ being the 5d Dirac spinor, whereas in 6d we use the Weyl spinor and define
\bea
\bar\eps &=& \eps^T c \otimes C \otimes C'
\eea
The difference is crucial, since $C_{11d}=c \otimes C \otimes C'$ is antisymmetric, while $C_{5d} = C\times C'$ is symmetric. So for example, we use this fact to show that 
\bea
\bar\eps \tau^{AB} \eps = (\eps^T C_{5d} \tau^{AB} \eps)^T = \eps^T (-C\tau^{AB}C^{-1}) C_{5d}^T \eps = - \bar\eps \tau^{AB} \eps
\eea
is vanishing. 

\section{Fierz identities}
Using 11d gamma matrices, we have for commuting $\eps$ subject to the 6d Weyl projection $\Gamma \eps = - \eps$
\bea
\eps\bar\eps &=& \frac{1}{16} \(\bar\eps\Gamma_M\eps\Gamma^M - \bar\eps\Gamma_{MA}\eps\Gamma^{MA}+\frac{1}{12}\Gamma_{MNPAB}\eps\Gamma^{MNPAB}\)P_+
\eea
where $P_+ = (1+\Gamma)/2$. Using 6d and 5d gamma matrices that we relate to the 11d gamma matrices through
\bea
\Gamma^M &=& \Gamma^M \otimes 1\cr
\Gamma^A &=& \Gamma \otimes \tau^A
\eea
where we recycle the same letter $\Gamma^M$ to refer to the 6d gamma matrices, we get
\bea
\eps\bar\eps &=& \frac{1}{16} \(\bar\eps\Gamma_M\eps\Gamma^M + \bar\eps\Gamma_M\tau^A\eps\Gamma^M\tau^A + \frac{1}{12} \bar\eps\Gamma_{MNP}\tau^{AB}\eps\Gamma^{MNP}\tau^{AB}\) P_+
\eea
We also have
\bea
\eps\bar\eta &=& \frac{1}{16} \(\bar\eta \eps + \bar\eta\tau^A\eps\tau^A - \frac{1}{2} \bar\eta\tau^{AB}\eps\tau^{AB}\)P_-\cr
&& + \frac{1}{16} \(-\frac{1}{2}\bar\eta\Gamma^{MN}\eps\Gamma_{MN}-\frac{1}{2}\bar\eta\Gamma^{MN}\tau^A\eps\Gamma_{MN}\tau^A + \frac{1}{4} \bar\eta\Gamma^{MN}\tau^{AB}\Gamma_{MN}\tau^{AB}\)P_-
\eea
and
\bea
\eta\bar\eps &=& -\frac{1}{16} \(\bar\eta\eps + \bar\eta\tau^A\eps\tau^A + \frac{1}{2} \bar\eta\tau^{AB}\eps\tau^{AB}\)P_+\cr
&& - \frac{1}{16} \(\frac{1}{2}\bar\eta\Gamma^{MN}\eps\Gamma_{MN} + \frac{1}{2}\bar\eta\Gamma^{MN}\tau^A\eps\Gamma_{MN}\tau^A + \frac{1}{4} \bar\eta\Gamma^{MN}\tau^{AB}\eps\Gamma_{MN}\tau^{AB}\)P_+
\eea
We have gamma matrix identities
\bea
\{\Gamma_{MN},\Gamma^{RST}\} &=& 12 \delta_{MN}^{RS} \Gamma^T + 2\Gamma_{MN}{}^{RST}\cr
\Gamma^{MNP}\Gamma^R \Gamma_{NP} &=& - 4 \Gamma^{MR} - 20 g^{MR}\cr
\Gamma^{MNP}\Gamma^{RST}\Gamma_{NP} &=& 4\Gamma^{MRST} + 12 g^{MR}\Gamma^{ST}\cr
\Gamma^{MNP}\Gamma^{RS}\Gamma_{MNP} &=& 24 \Gamma^{RS}\cr
\tau^A\tau^B\tau^A &=& - 3\tau^B\cr
\tau^A\tau^{BC}\tau^A &=& \tau^{BC}
\eea
These Fierz identities can be used to obtain the closure relation for the abelian $(2,0)$ 6d tensor multiplet on the fermion. From (\ref{6dtensor}) we first get
\bea
\delta^2 \psi &=& i \(\frac{1}{4} \Gamma^{MNP} \eps\bar\eps\Gamma_{NP} - \Gamma^M \tau^A \eps\bar\eps \tau^A\) D_M\psi\cr
&& + i \(-\frac{1}{4}\Gamma^{MNP}\eps\bar\eta\Gamma_{MNP}+\Gamma^M\tau^A\eps\bar\eta\tau^A\Gamma_M - 4\tau^A\eta\bar\eps\tau^A\)\psi
\eea
We get from the first line
\bea
(\delta^2 \psi)_{1st} &=& - i S^M D_M \psi \cr
&& + \frac{3i}{8} S_R \Gamma^R \Gamma^M D_M \psi\cr
&& - \frac{1}{8} \bar\eps\Gamma_R\tau^A\eps \tau^A\Gamma^R\Gamma^M D_M\psi
\eea
This is the complete result, the contribution from $\Gamma^{RST} \tau^{CD}$ completely cancels out by some lucky circumstances. Let us now move on to the second line,
\bea
(\delta^2 \psi)_{2nd} &=& \frac{i}{2} \bar\eta \Gamma_{MN}\eps \Gamma^{MN} \psi\cr
&=& - \frac{i}{4} D_M S_N \Gamma^{MN} \psi
\eea
This term combines with a term from the first line into a full Lie derivative
\bea
- i S^M D_M \psi - \frac{i}{4} D_M S_N \Gamma^{MN} \psi &=:& - i \L_S \psi
\eea
and so we get one of the closure relations presented in the main text. 

Reduction to $(1,0)$ supersymmetry is done by imposing the condition
\bea
\tau^5 \eps = - \eps
\eea
and then we obtain the Fierz identity
\bea
\eps \bar\eps &=& \frac{1}{8} \(\bar\eps \eps + \bar\eps \gamma_m \eps \gamma^m\) - \frac{1}{64} \bar\eps\gamma_{mn}\tau^{ij}\eps \gamma^{mn} \tau^{ij}
\eea
We also have made use of the following 5d gamma matrix identities
\bea
\gamma^m \gamma^p \gamma_m &=& - 3\gamma^p\cr
\gamma^m \gamma^{pq} \gamma_m &=& \gamma^{pq}\cr
\gamma^{mn} \gamma^p \gamma_{mn} &=& - 4 \gamma^p\cr
\gamma^{mn} \gamma^{pq} \gamma_{mn} &=& 4 \gamma^{pq}
\eea

\section{Selfduality from a Lagrangian}
Here we review the main result of \cite{Lee:2000kc}, \cite{Bonetti:2012fn}. For abelian case, we have the selfduality equation 
\bea
\frac{1}{r}\F_{mn} &=& - \frac{1}{6} \E_{mn}{}^{qrs} H_{qrs} + \frac{1}{2} \E_{mn}{}^{qrs} \F_{qr} \kappa_s
\eea
where 
\bea
\F_{mn} &=& F_{mn} + \partial_0 B_{mn}\cr
F_{mn} &=& 2\partial_m A_n
\eea
By using the Bianchi identity for $H_{qrs}$ we get the Maxwell equation
\bea
D^m \(\frac{1}{r} \F_{mn}\) - \frac{1}{2} \E_{mn}{}^{qrs} D^m\(\F_{qr} \kappa_s\) &=& 0
\eea
which follows from an action
\bea
\L &=& \frac{1}{4r} \F^{mn} \F_{mn} - \frac{1}{8} \E^{mnpqr} \F_{mn} \F_{pq} \kappa_r + \frac{1}{24} \E^{mnpqr} B_{mn} \partial_0 H_{rpq}
\eea
by varying $A_m$. But now we can also vary $B_{mn}$ and then we get
\bea
\partial_0\(\frac{1}{2r}\F_{mn} - \frac{1}{4} \E_{mn}{}^{pqr} \F_{pq}\kappa_r - \frac{1}{12} \E_{mn}{}^{pqr} H_{pqr}\) &=& 0
\eea
The last term can be replaced by 
\bea
-\frac{1}{24} \E^{mnpqr} B_{mn} \partial_0 H_{rpq} =  \frac{1}{8} \E^{mnpqr}  \F_{mn} \partial_0^{-1}\partial_r \F_{pq}
\eea
where we note that $\partial_0 H_{rpq} = 3 \partial_r \F_{pq}$. Thus we can write 
\bea
\L &=&  \frac{1}{4r} \F^{mn} \F_{mn} + \frac{1}{8} \E^{mnpqr} \F_{mn} \partial_0^{-1} \(\partial_r - \kappa_r \partial_0\) \F_{pq}
\eea

\end{document}